\documentclass[useAMS,usenatbib]{mn2e} 
\usepackage{aas_macros}
\usepackage{graphics}
\usepackage[pdftex]{graphicx}
\usepackage{epstopdf}
\usepackage{epsfig}  
\usepackage{natbib} 
\usepackage{dingbat}
\usepackage{float}
\usepackage{amsmath}
\usepackage{times}
\usepackage[varg]{txfonts}
\usepackage{verbatim} 
\usepackage{multirow,bigdelim} 
\usepackage{color}
\usepackage{vmargin}
\setmarginsrb{1.2cm}{1cm}{1.2cm}{1cm}{1cm}{1cm}{1cm}{1cm}

\newcommand{\hmsun}{{\, h^{-1}\rm~M}_\odot}

  \makeatletter
    \renewcommand{\paragraph}{\@startsection{paragraph}{4}{\z@}%
      {-3.25ex\@plus -1ex \@minus -.2ex}%
      {1.5ex \@plus .2ex}%
      {\normalfont\small\centering}}
     
    \renewcommand{\subparagraph}{\@startsection{subparagraph}{5}{\z@}%
      {-3.25ex\@plus -1ex \@minus -.2ex}%
      {1.5ex \@plus .2ex}%
      {\normalfont\small\centering}}
    \makeatother

\newcommand{\ginnungagap}{{\sc Ginnungagap}}

\newcommand{\gadget}{{\sc Gadget}}

\newcommand{\hMpc}{{ \textit{h}$^{-1}$~Mpc}}

\title[Simulated clusters]{Galaxy clusters in simulations of the local Universe: a matter of constraints}
\author[Sorce \& Tempel]
{{Jenny G. Sorce$^{1,2,3}$\thanks{E-mail: \text{jenny.sorce@univ-lyon1.fr / jsorce@aip.de}}, 
Elmo Tempel$^{3,4}$
}\\
$^1$Univ Lyon, Univ Lyon1, Ens de Lyon, CNRS, Centre de Recherche Astrophysique de Lyon UMR5574, F-69230, Saint-Genis-Laval, France\\
$^2$Universit\'e de Strasbourg, CNRS, Observatoire astronomique de Strasbourg, UMR 7550, F-67000 Strasbourg, France\\
$^3$Leibniz-Institut f\"{u}r Astrophysik Postdam (AIP), An der Sternwarte 16, 14482 Potsdam, Germany\\
$^4$Tartu Observatory, University of Tartu, Observatooriumi 1, 61602 T\~oravere, Estonia
}

\begin{document}

\date{}

\pagerange{\pageref{firstpage}--\pageref{lastpage}} \pubyear{2017}

\maketitle

\label{firstpage}

\begin{abstract}
\indent 
To study the full formation and evolution history of galaxy clusters and their population, high resolution simulations of the latter are flourishing. However comparing observed clusters to the simulated ones on a one-to-one basis to refine the models and theories down to the details is non trivial. The large variety of clusters limits the comparisons between observed and numerical clusters. Simulations resembling the local Universe down to the cluster scales permit pushing the limit. Simulated and observed clusters can be matched on a one-to-one basis for direct comparisons provided that clusters are well reproduced besides being in the proper large scale environment. Comparing random and local-Universe like simulations obtained with differently grouped observational catalogs of peculiar velocities, this paper shows that the grouping scheme used to remove non-linear motions in the catalogs that constrain the simulations affects the quality of the numerical clusters. With \emph{a less aggressive grouping scheme - galaxies still falling onto clusters are preserved - combined with a bias minimization scheme}, the mass of the dark matter halos, simulacra for 5 local clusters - Virgo, Centaurus, Coma, Hydra and Perseus - is increased by 39\% closing the gap with observational mass estimates. Simulacra are found on average in 89\% of the simulations, an increase of 5\% with respect to the previous grouping scheme. The only exception is Perseus. Since the Perseus-Pisces region is not well covered by the used peculiar velocity catalog, the latest release let us foresee a better simulacrum for Perseus in a near future.
\end{abstract}


\section{Introduction}

Clusters of galaxies are excellent cosmological probes. Understanding their formation and evolution is thus an entirely logical step in our quest towards understanding the Universe as a whole. However, accessing detailed information about galaxy clusters via observations is far from direct and the extracted information might suffer from observational biases. To complement observational studies, high resolution simulations of galaxies clusters are now flourishing \citep[e.g. see][for a non-extensive list]{2013ApJ...763...70W,2013ApJ...767...23W,2015MNRAS.452.1982W,2016MNRAS.459.4408M,2016MNRAS.457.4063S,2016MNRAS.458.1096E,2016MNRAS.458.4052C,2016MNRAS.459.2973S,2017MNRAS.464.2027A,2017MNRAS.465.2584B,2017MNRAS.465..213B,2017MNRAS.tmp..205H} and comparisons between observed and simulated clusters emerge \citep[e.g.][]{2017PASJ...69...14S,2017MNRAS.466.2658J,2017MNRAS.468.1962N}. Still these comparisons are limited because of the large variety of cluster types in terms of morphology, mass, evolution stage, etc \citep{1988S&T....75...16S}. Selecting adequately the simulated cluster candidates to be compared with a given observed cluster is not immediately obvious and in a certain sense it is almost impossible to find the exact counterpart \citep{2015ApJ...807...88G}. Determining how effectively a numerical cluster represents an observed one is actually subject to uncertainties. 

One way to reduce these uncertainties is to use simulations that resemble a portion of the Universe with well observed clusters. Such simulations host clusters similar to the observed ones in the proper environment and thus make the comparisons between observations and simulations even more legitimate. The most well observed part of the Universe is undeniably the local Universe. Consequently, efficient simulations of the local Universe down to the cluster scales constitute the optimal choice to perform the detailed comparisons between observations and simulations mentioned earlier. Such simulations unlike typical ones stem from a set of constraints in addition to abiding to a cosmological prior \citep{1987ApJ...323L.103B,2010arXiv1005.2687G,2010MNRAS.406.1007L,2013MNRAS.429L..84K}. These constraints can be either peculiar velocities \citep[e.g][]{2003ApJ...596...19K} equivalently distances \citep[e.g.][]{2016MNRAS.457..172L} or redshift surveys \citep[e.g.][]{1989ApJ...336L...5B,1990ApJ...364..370B,2009MNRAS.400..183K,2013MNRAS.435.2065H,2016ApJ...831..164W}. The initial conditions constrained by the aforementioned measurements can be produced either forward \citep[e.g][]{2008MNRAS.389..497K,2013MNRAS.432..894J,2013ApJ...772...63W,2014ApJ...794...94W} or backwards \citep[e.g.][]{1990ApJ...364..349D,1999ApJ...520..413Z,1993ApJ...415L...5G,2008MNRAS.383.1292L}. We use the latter in this paper and in our previous papers. Our first simulations resembling the local Universe obtained successively with the first catalog \citep{2008ApJ...676..184T} of peculiar velocities of the Cosmicflows project and the second one \citep{2013AJ....146...86T} hosted the local Large Scale Structure with a remarkable accuracy \citep{2014MNRAS.437.3586S,2016MNRAS.455.2078S}. In addition, large overdensities at the location of prominent local clusters clearly appeared in these simulations. However, the application of a halo finder to these simulations revealed that clusters, our closest neighbor Virgo cluster excluded \citep{2016MNRAS.460.2015S}, are not strongly reproduced in the simulation:  
the largest object in the large overdensity is not massive enough with respect to expectations based on observational estimates. In a companion paper  \citep{2017MNRAS.469.2859S}, we showed that the grouping technique applied upstream on the catalog of constraints used to build the initial conditions might affect the production of massive clusters although it does not affect overall the local Large Scale Structure. Namely the overdense regions hosting the clusters are more or less pronounced depending on the grouping scheme used.

The grouping applied to the constraints is an absolute requirement. Indeed the technique we use to build the initial conditions is linear but constraints include galaxies and their velocities in various environments including dense environments like clusters. Therefore, non-linear motions find their way into the catalog of constraints and are passed along to the reconstruction technique that cannot handle them in an appropriate manner. This same phenomenon is visible in redshift surveys in the form of fingers of god \citep{1972MNRAS.156P...1J} and Kaiser's effect \citep{1987MNRAS.227....1K}. These effects need to be suppressed since they affect reconstructions based on redshift surveys \citep{2012MNRAS.427L..35K}. A different modeling can be applied to the small scales with respect to the large scales: Spherical collapse versus Lagrangian perturbation \citep{2013MNRAS.435.2065H}. However, a certain balance is necessary: while it is necessary to group to suppress non-linear motions, galaxies in the field and a fortiori galaxies infalling onto clusters are essential to retrieve the proper density field and obtain an optimal reconstruction of the local Universe \citep{2017MNRAS.468.1812S}. In this paper, we show that as expected what is true for the reconstruction of the local Universe \citep{2017MNRAS.469.2859S} is also valid for its simulations.

This paper starts with a brief description of the catalog of constraints, the grouping algorithm applied to it and the different steps to build the constrained initial conditions. Then, the resulting  simulations of the local Universe are analyzed and compared to those obtained with the earlier released version of the grouped catalog as well as to random simulations. Finally, a conclusion closes the paper.   


\section{Building constrained initial conditions}

The different steps to produce constrained initial conditions used in the project have been widely described and summarized in previous papers \citep[e.g.][]{2016MNRAS.455.2078S}. In the following they are briefly reminded. 

\subsection{The Catalog}

The second catalog of radial peculiar velocities or more precisely of direct distance measurements of the Cosmicflows project constitutes our set of constraints.  Published in \citet{2013AJ....146...86T}, it contains more than 8,000 accurate galaxy distances  mostly ($\sim88\%$) obtained with the Tully-Fisher \citep{1977A&A....54..661T} and the Fundamental Plane \citep{2001MNRAS.321..277C} methods. Cepheids \citep{2001ApJ...553...47F}, Tip of the Red Giant Branch \citep{1993ApJ...417..553L}, Surface Brightness Fluctuation \citep{2001ApJ...546..681T}, supernovae of type Ia \citep{2007ApJ...659..122J} and other miscellaneous methods constitute the remaining $\sim12\%$. It extends up to about 250~\hMpc\ and about 50\% of the data are within 70~\hMpc\ and 90\% within 160~\hMpc. 

\subsection{The Grouping Scheme}
The grouping scheme is widely described in \citet{2016A&A...588A..14T}  and our application to the catalog of constraints is detailed in  \citet{2017MNRAS.469.2859S}. A brief description is given here as a reminder.

 \citet{2016A&A...588A..14T} introduced a new grouping method (hereafter Tempel grouping scheme). This method is based on a widely used Friends of Friends (FoF) percolation method, where different linking lengths in radial (along the line of sight) and in transversal (in the plane of the sky) directions are used but the conventional FoF groups are refined using multimodality analysis. More precisely, \citet{2016A&A...588A..14T} use a model-based clustering analysis to check the multimodality of groups found by the FoF algorithm and they separate nearby/merging systems. In \citet{2017MNRAS.469.2859S}, we tested different linking lengths and settled for the default one (0.25~\hMpc\ at redshift zero) so as to group sufficiently to remove non-linear motions without large residuals and not too much so as to preserve the infall onto the clusters.
 
The grouping scheme thus provides the groups to which the different galaxies that populate the second catalog of Cosmicflows belong to as well as their total velocity. This information is combined with the galaxy distance estimates given by the second catalog of Cosmicflows to access galaxy radial peculiar velocities (the constraints).

Furthermore, the constrained simulations obtained with this grouping scheme are to be compared with the first generation of constrained simulations obtained with the second Cosmicflows catalog of radial peculiar velocities and the grouping version (hereafter Tully grouping scheme) released via the Extragalactic Distance Database\footnote{http://edd.ifa.hawaii.edu/} \citep{2009AJ....138..323T}. We remind that this earlier scheme is based on literature groups and thus is not a systematic scheme: within 30~Mpc, groups are those identified by \citet{1987ApJ...321..280T}, further away groups are those given in the literature like Abell's catalog \citep{1989ApJS...70....1A}. 

\subsection{Bias minimization, Reconstruction, Reverse Zel'dovich Approximation, Constrained Realizations and Rescaling}

Five more steps are required to complete the construction of the constrained initial conditions:
\begin{enumerate}
\item Minimization of the biases \citep{2015MNRAS.450.2644S} inherent to any observational radial peculiar velocity catalog. This minimization permits removing the spurious infall onto the local Volume and gives a proper Virgo cluster in the simulations and larger masses for the other nearby clusters.
\item Reconstruction of the cosmic displacement field with the Wiener-Filter (WF) technique \citep[linear minimum variance estimator, in abridged form WF,][]{1995ApJ...449..446Z,1999ApJ...520..413Z} applied to the peculiar velocity constraints.
\item Relocation of the constraints to the positions of their progenitors using the Reverse Zel'dovich Approximation and the reconstructed cosmic displacement field \citep{2013MNRAS.430..888D} and replacing noisy radial peculiar velocities by their WF 3D reconstructions \citep{2014MNRAS.437.3586S} to ensure that structures are at the proper position at redshift zero.
\item Production of the density fields constrained by the modified observational peculiar velocities combined with a random realization to restore statistically the missing structures using the Constrained Realization technique \citep[CR,][]{1991ApJ...380L...5H,1992ApJ...384..448H,1996MNRAS.281...84V}. 
\item Rescaling of the density fields to build constrained initial conditions and increasing the resolution by adding small scale features (e.g. \ginnungagap\ code\footnote{https://github.com/ginnungagapgroup/ginnungagap}).
\end{enumerate}

To enrich our comparisons a set of random (typical) initial conditions is prepared.
All the initial conditions are built within the Planck cosmology framework \citep[$\Omega_m$=0.307, $\Omega_\Lambda$=0.693, H$_0$=67.77, $\sigma_8$~=~0.829,][]{2014A&A...571A..16P} in 500 \hMpc\ boxes with 512$^3$ particles (particle mass: 8$\times$10$^{10}$~$\hmsun$). The simulations are run with the N-body code \gadget\ \citep[][]{2005MNRAS.364.1105S}.

\section{local Universe like simulations}

\begin{figure*}
\vspace{-7cm}
\includegraphics[width=0.8 \textwidth]{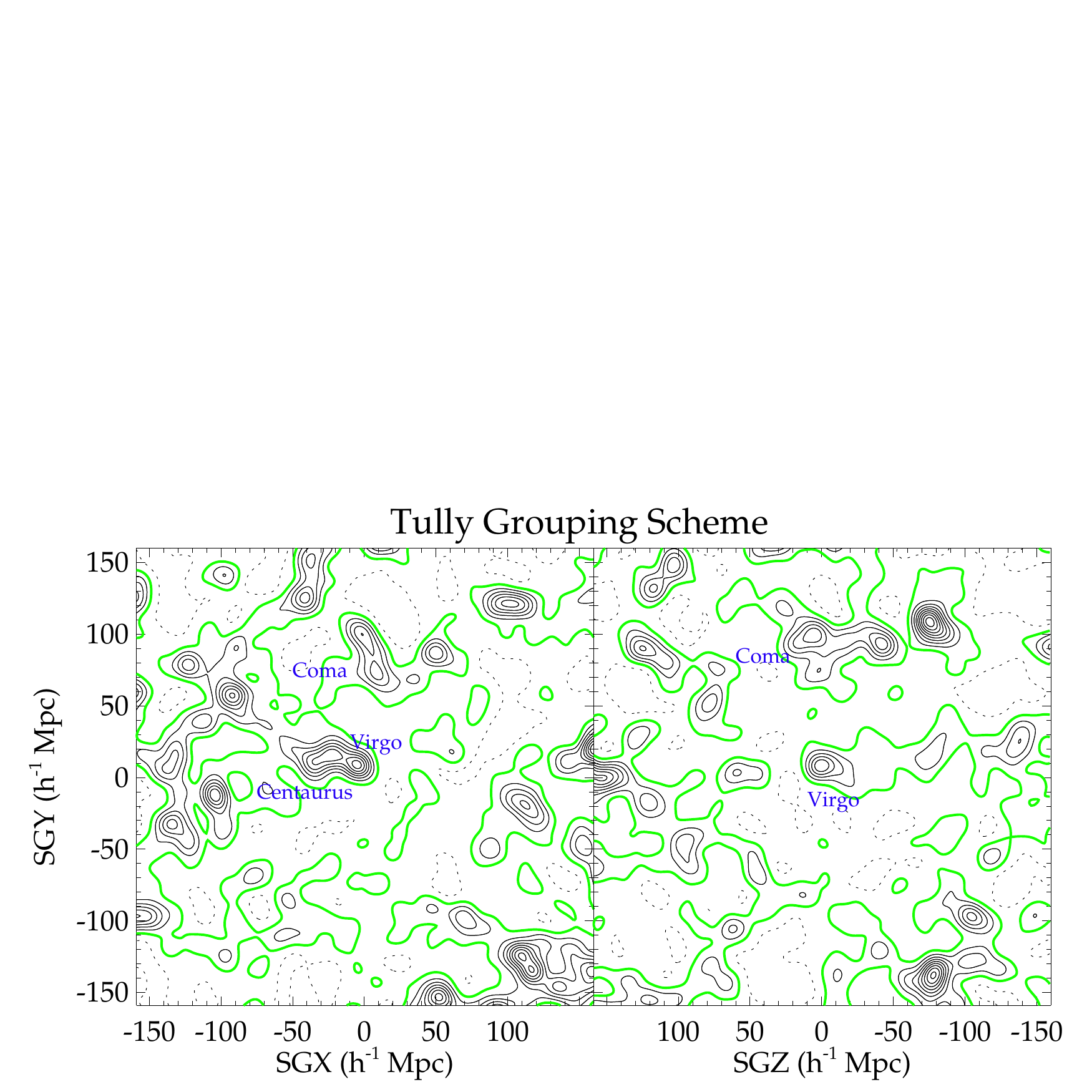}\\
\vspace{-7cm}
\includegraphics[width=0.8 \textwidth]{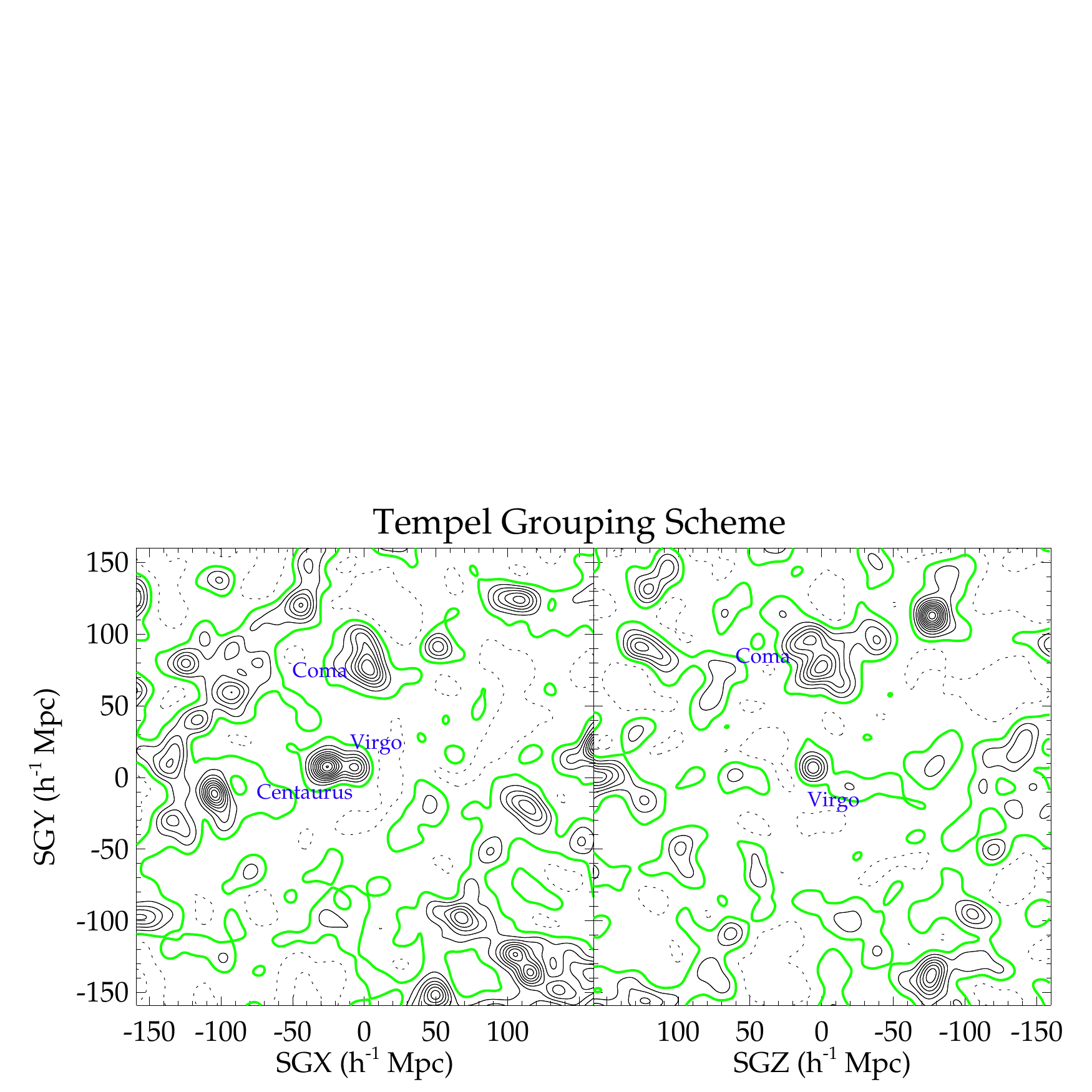} 
\caption{Supergalactic slices of the local Large Scale Structure obtained in constrained simulations. A 5~\hMpc\ smoothing scale has been applied to the fields. Different grouping schemes are used to remove non-linear motions from the constraint-catalog of galaxy radial peculiar velocities. Top: Tully grouping scheme, released with the catalog and used for the first generation of constrained simulations based on the second catalog of Cosmicflows. Bottom: Tempel grouping scheme, tested in this paper. The contours stand for the density. Solid lines show overdensities while dotted lines represent underdensities. The green color is the mean field. A few structures are named in blue. Overall the local Large Scale Structure is properly reproduced in both simulations. The differences appear only at the cluster scale. For instance, Coma and Centaurus are more clearly defined in the simulation obtained with Tempel grouping scheme.}
\label{fig:LSS}
\end{figure*}

In this section, nine simulations of each type (Tully grouping scheme, Tempel grouping scheme and random) are used for further comparisons. A total of 27 simulations is thus run and every group of 3 simulations (1 simulation per type in a given group) is based on the exact same random realization (cf. CR). For each type, nine simulations permit studying the stability of constrained simulations and the same random seeds for each group allows us to study the effect of the grouping technique. First the local Large Scale Structure is the object of focus before directing the efforts towards studying the clusters (dark matter halos). For simplification, throughout the rest of the paper, ``Tully and Tempel constrained simulations'' is adopted as a shortened notation.

\subsection{The local Large Scale Structure}
\begin{figure*}
\vspace{-1cm}
\includegraphics[width=1 \textwidth]{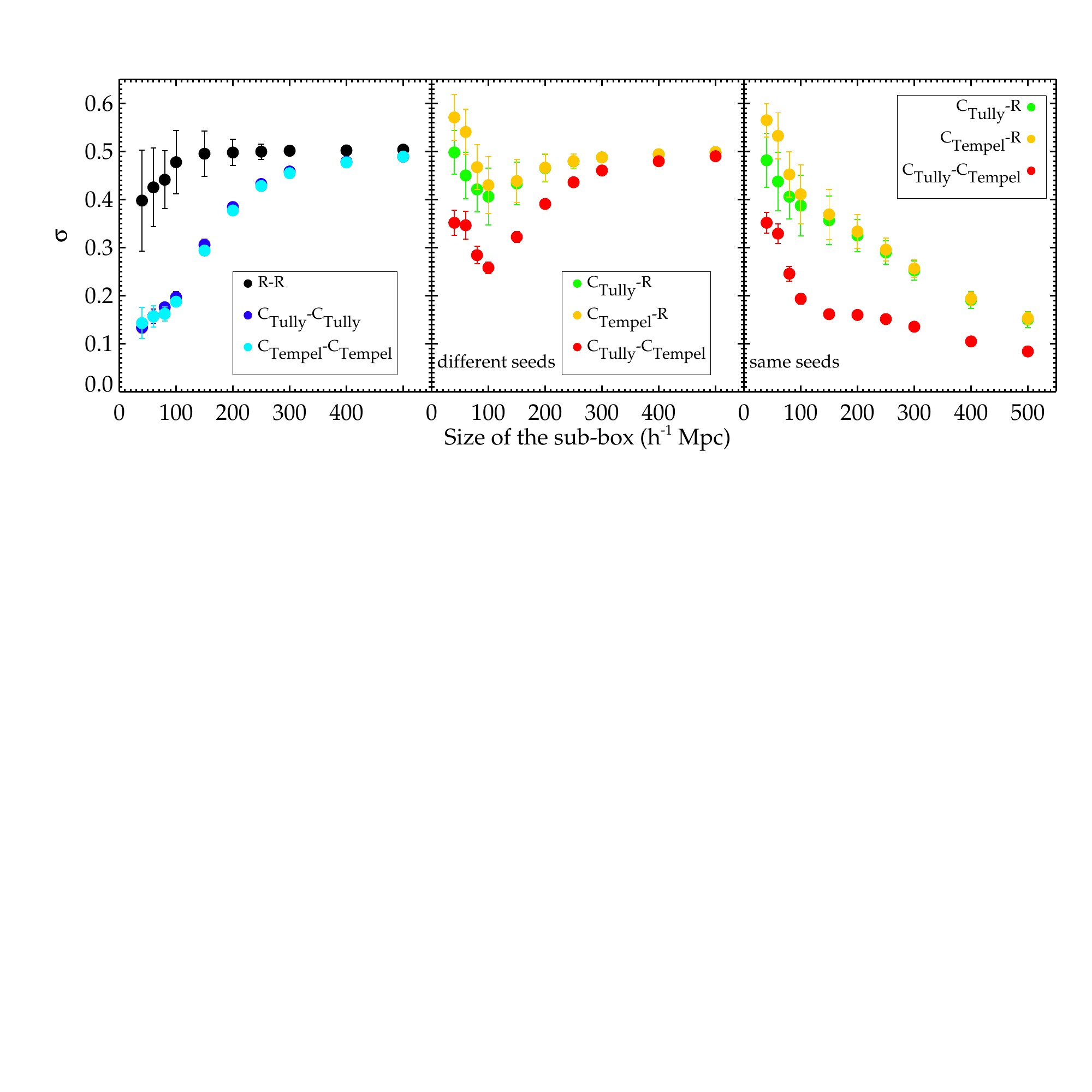}
\vspace{-11cm}
\caption{Average variance (filled circle) and its standard deviation (error bar) between density fields of simulations as a function of the size of the compared sub-box. From left to right: comparisons of pairs of random (R, black) and constrained (C$_{\rm Tully}$ dark blue, C$_{\rm Tempel}$ light blue) simulations, comparisons between random and constrained simulations (green and yellow) as well as between constrained simulations obtained with different grouping schemes (red) that do not share the same random realization (middle) and that share the same random realization (right). The goal of the constrained simulations is fulfilled: the cosmic variance is reduced with respect to that of random simulations.}
\label{fig:cosvar}
\end{figure*}

\begin{figure}
\vspace{-2cm}

\includegraphics[width=0.5 \textwidth]{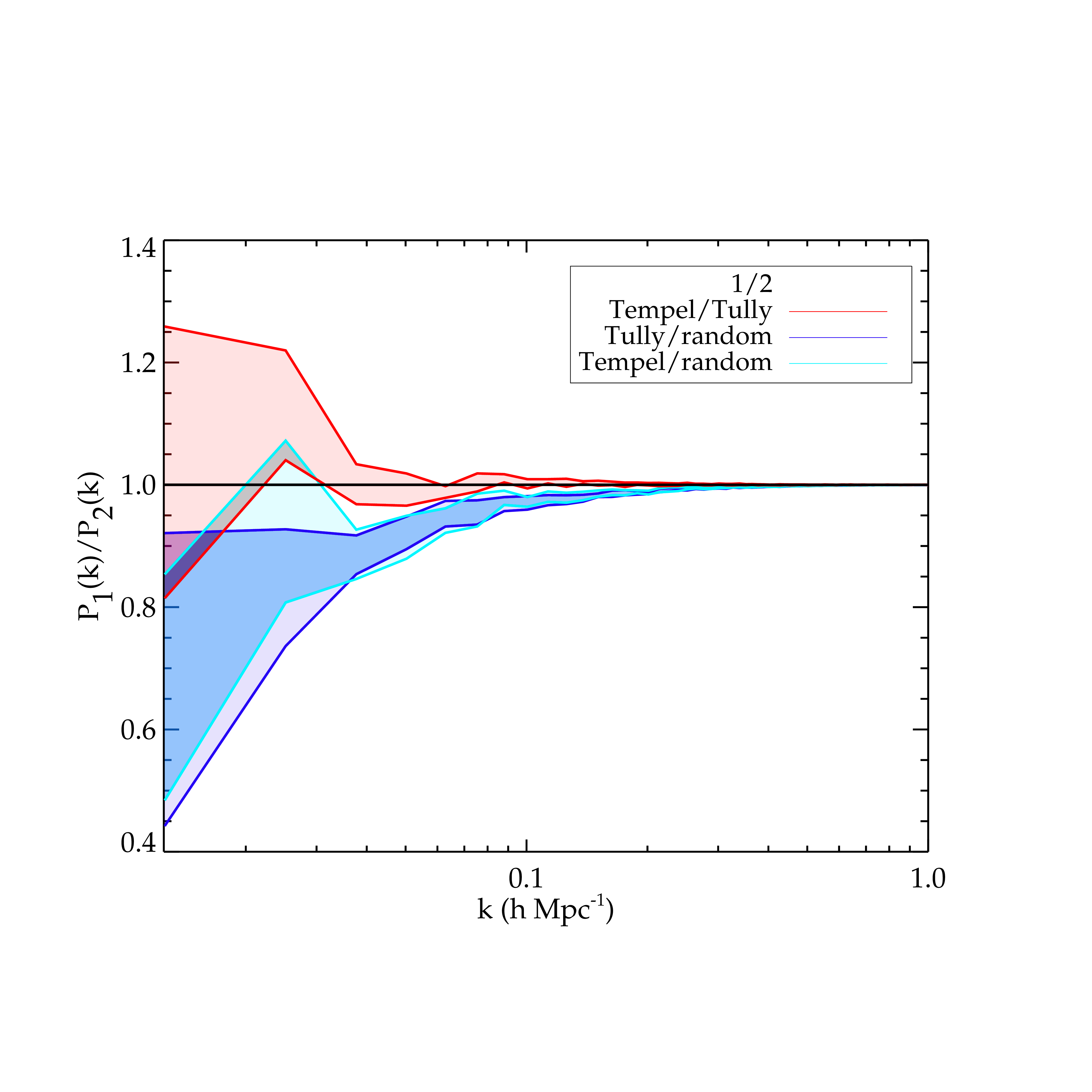}
\vspace{-3.2cm}

\includegraphics[width=0.5 \textwidth]{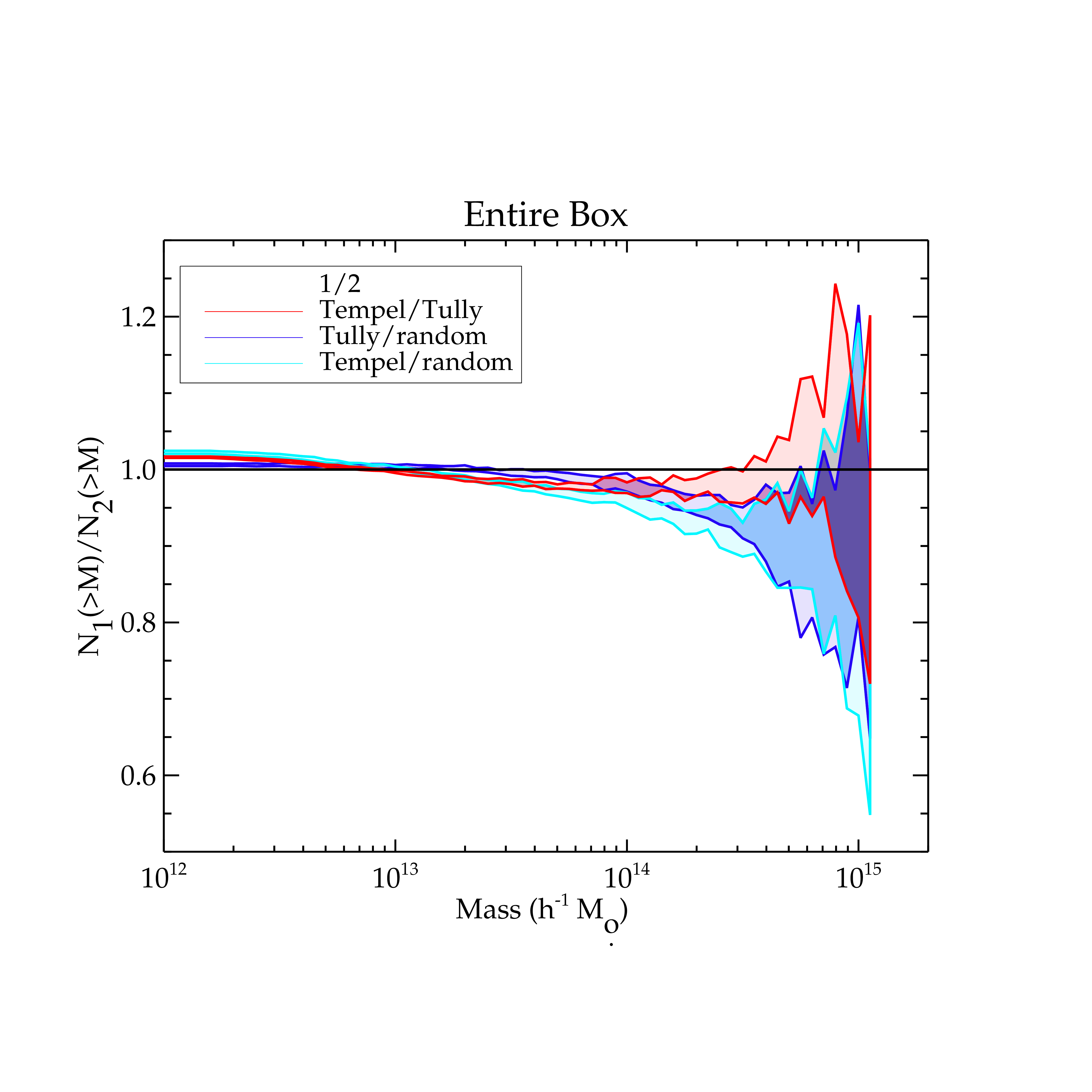}

\vspace{-2.8cm}
\includegraphics[width=0.5 \textwidth]{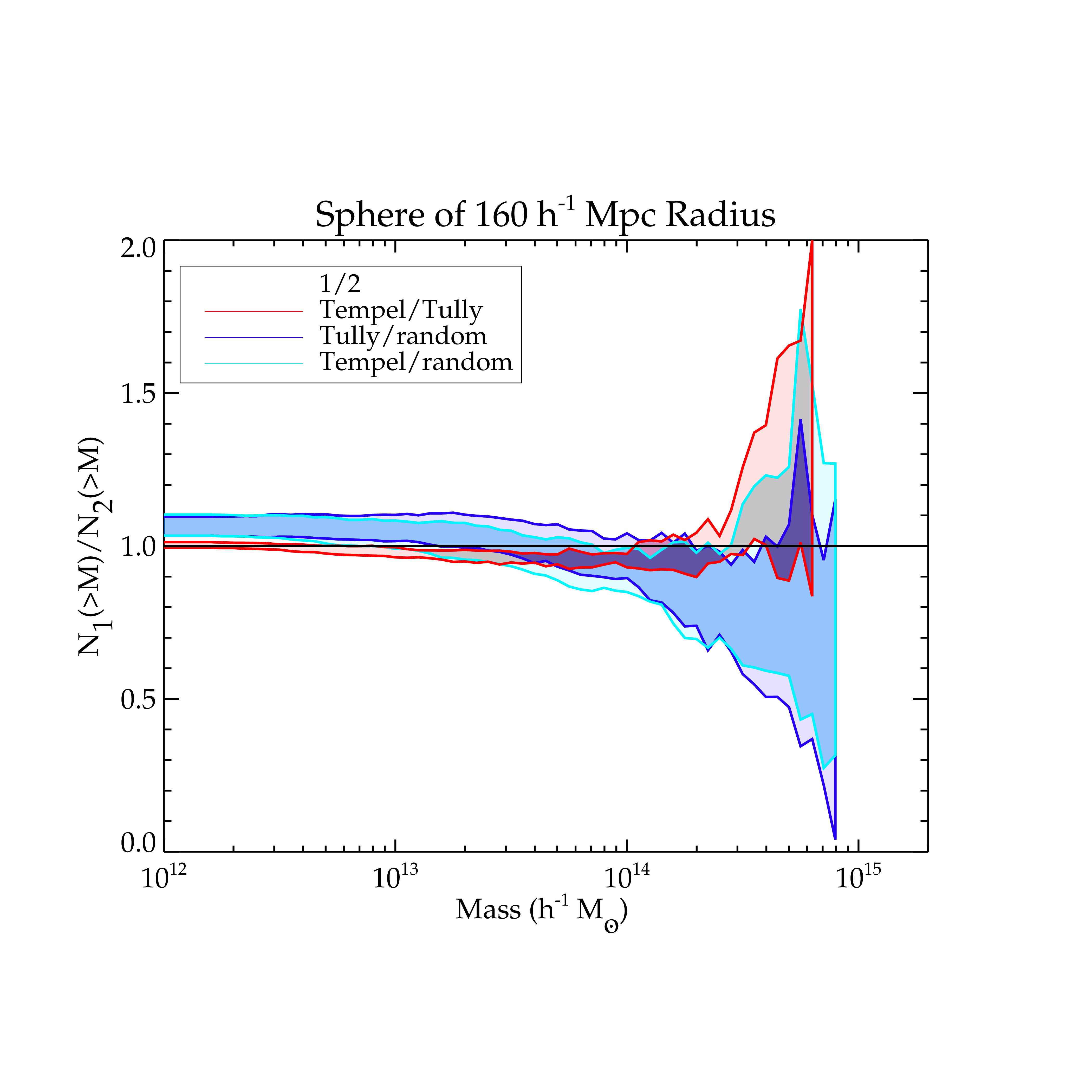}
\vspace{-2cm}

\caption{Top: 1$\sigma$ confidence interval of the ratio of the power spectra of constrained and random simulations (blue) and of constrained simulations (red). Middle and bottom: the same as the top panel but for the mass functions of the entire box and of a 160 \hMpc\ radius sphere.}
\label{fig:powspecmass}
\end{figure}

Fig. \ref{fig:LSS} shows the local Large Scale Structure obtained for two constrained simulations. The top panel presents two supergalactic slices of the local Universe obtained with Tully grouped version of the catalog while the bottom panel gives the local structures resulting from Tempel grouping scheme. The solid contours stand for the overdensities while the dotted ones represent the underdensities. The green color stands for the mean field. A few structures are identified with blue names. Overall the local Large Scale Structure is well reproduced in both cases. It is very similar and the differences appear only in the details, at the cluster scale level. For instance, Coma appears more distinctly with Tempel grouping scheme than with Tully's. In addition, if Virgo is well delimited in both cases, Centaurus appears more clearly for Tempel grouping scheme than for Tully's.\\

\citet{2016MNRAS.455.2078S} showed that the cosmic variance is reduced by a factor 2 to 3 in the inner part of the box for the first generation of constrained simulations based on the second catalog of Cosmicflows and Tully grouping scheme. It is interesting to quantify it when changing the grouping scheme for Tempel's. Fig. \ref{fig:cosvar} shows the average of the variances (filled circles) and their standard deviation (error bars) obtained when comparing pairs of random (R) and pairs of constrained (C$_{\rm Tully}$, C$_{\rm Tempel}$) simulations. 

The variance is defined as the scatter around the 1:1 relation obtained when comparing every cell from a simulation to its exact counterpart in the other simulation of the pair. Once all variances are derived their mean and standard deviation are derived and plotted as a filled circle with error bars. Since most of the constraints are within the inner part of the box, it is reasonable to compare not only the entire boxes but also their inner parts. Consequently the process is repeated cutting the boxes to compare smaller and smaller regions. 

The first panel of Fig. \ref{fig:cosvar} shows the variance between pairs of simulations of the same nature. Clearly and as expected the random simulations differ on average by $\sim$32\% more from each others (black) than the constrained ones (blue). An identical reduction of the cosmic variance by a factor 1.5 on average and 2.5-3 within the inner part of the box is observed for both grouping schemes. There is a limit to the method used to derive the cosmic variance. This is visible when comparing only the inner, and thus smaller, parts of the boxes: the mean variance decreases by 20\% for the pairs of random simulations. This is entirely due to the higher probability of finding small empty regions than large empty regions. Note however that since it is not improbable to find dense regions (even if the probability is low), the standard deviation is on average about 10 times larger when considering the inner part of the random box than when considering the entire random box. This inconvenience happens only for the pairs of random simulations. It is indeed well known that there are structures in the inner part of the box for the constrained simulations by construction (the local Universe has structures there).

The second and third panels of Fig. \ref{fig:cosvar} show the variance between pairs of random and constrained simulations as well as between pairs of constrained simulations obtained with different grouping schemes. The third panel averages only the variances obtained for  pairs of simulations sharing the same random realization, while the second panel averages the variances of the other pairs (not sharing the same random realization). Four points are worth noticing:
\begin{enumerate}
\item The average variance obtained for pairs of constrained simulations based on different grouping schemes is smaller by about 20\% than that obtained for pairs of random and constrained simulations. In addition when comparing large volumes, the shape of the curve drawn by the variances is identical to that obtained when comparing constrained simulations obtained with the same grouping scheme. This confirms that overall the grouping scheme does not affect the simulation of the local Large Scale Structure. 
\item However when reaching the inner part of the box, the mean variance between the simulations increases by up to 20-25\% with respect to its minimum rather than continuing its decrease. While the result is completely expected when comparing random and constrained simulations - higher probability of finding a small empty region in the random simulation to be compared to the known structures in the very nearby Universe - in the comparison between constrained simulations based on different grouping schemes, the finding is entirely due to the small differences noted at the cluster scale in Fig. \ref{fig:LSS}.
\item In the third panel, if the smallest by 40-45\% average variance is still that obtained when comparing constrained simulations, it is worth noticing that the mean variance increases by a factor up to 3-4 with the decrease in size of the compared regions. While this is entirely due to the fact that the weakly constrained part of the box - hence the random realization - dominates to a large extent when comparing the totality of the random and constrained boxes, when comparing constrained simulations it emphasizes that the Large Scale Structure is quite unaffected by the grouping scheme down to volumes of $\sim$(100~\hMpc)$^3$. The latter affects the simulations only at the cluster scale. The shape of the curve drawn by the variances when comparing only the inner part of the box is indeed similar when comparing only constrained simulations sharing the same random realization and when comparing constrained simulations whatever random realization they have been constructed of.
\item  Within the inner part of the box, Tempel grouping scheme results in constrained simulations that differ by 20\% more from the random simulations than those obtained with Tully grouping scheme. This is in agreement with the results found in \citet{2017MNRAS.469.2859S}: the densities are more pronounced with Tempel grouping scheme than with Tully's, hence the constrained simulations differ more from the random ones in the former case than in the latter: a majority of underdensities are compared with higher overdensities.\\
\end{enumerate}

Before focusing on the clusters (dark matter halos) in a detailed way in the two different types of constrained simulations, it is worth comparing the power spectra and mass functions of the simulations. Fig. \ref{fig:powspecmass} shows the 1$\sigma$ confidence interval of the ratios of the power spectra and mass functions of the entire box and for a 160~\hMpc\ radius sphere of constrained and random simulations. The Amiga's halo finder is used to find the dark matter halos in all the simulations \citep{2009ApJS..182..608K}.

Overall the power spectra of the constrained simulations are below those of the random simulations on large scales as already noticed by \citet{2016MNRAS.455.2078S}. Tests conducted on mock catalogs are in favor of the data as the most likely culprit, either as an intrinsic property or/and because of their modeling via for instance their grouping (rather than the succession of well-established mathematical procedures including Wiener filtering, Reverse Zel'dovich Approximation and Constrained Realizations). This paper focuses on studying the impact of the data grouping modeling. Indeed before any possibility of concluding that this behavior is an intrinsic property of our local environment, any data modeling must be investigated. The power spectra of the simulations obtained with Tempel grouping scheme have on average slightly higher (10\%) values than those of the simulations obtained with Tully grouping scheme on large scales: the light blue zone tends to be above the dark blue zone. Consequently, the grouping scheme is partly responsible for the observation made by \citet{2016MNRAS.455.2078S}. Namely the power spectra of the constrained simulations obtained with Tempel grouping scheme have smaller values than those of the random simulations but to a lesser extent than those obtained with Tully grouping scheme; alternatively the red zone is above 1.0 meaning that the power spectra of Tempel constrained simulations have higher values than those of Tully constrained simulations on the large scales. This improves the probability of the local power spectrum given the Planck power spectrum. 

Regarding the mass functions in the entire box or in the sphere, constrained simulations tend to have less massive halos than random simulations as already observed by \citet{2016MNRAS.455.2078S}. The same discussion as above is here also valid. Tempel constrained simulations have on average 1.5 more massive (above 2$\times$10$^{14}~\hmsun$) halos than Tully constrained simulations within the 160 \hMpc\ radius sphere as shown by the light blue zone that is on average above the dark blue one or by the red zone that is clearly above 1.0 on average at the high mass end. These observations reinforce our expectations that Tempel scheme produces constrained simulations with more massive halos than Tully scheme. 

Although the grouping scheme alleviates the tension between power spectra and mass functions of constrained and random simulations, further investigations (an investigation of the data uncertainty modeling is currently underway) are necessary to conclude as to the reason for the observed residual. It could be either an intrinsic property of our local environment or another data modeling that needs improvement or both. Still, the next section proves that the constrained simulations are completely valid at least within 30~\hMpc\ where Virgo, Hydra and Centaurus clusters are perfectly reproduced as well as for zoom-in simulations of these clusters.

\subsection{Local clusters of galaxies}

\begin{table*}
\begin{center} 
\begin{tabular}{lrrrrrrrrl}
\hline \hline
& \multicolumn{5}{c}{Observation} & & \multicolumn{2}{c}{Simulation}\\
(1) & (2) & (3) & (4) & (5) & (6) & & (7) & (8)\\
Cluster & sgl & sgb & d & M &M$_{200}$& & M$_{200}$ &  M$_{200}$  \\
 & ($^\circ$) & ($^\circ$) & (Mpc) &(10$^{14}$~M$_\odot$) & (10$^{14}~\hmsun$) && (10$^{14}~\hmsun$)&  (10$^{14}~\hmsun$) \\
\hline
Virgo 	& 103.0008 & -2.3248  &14.9 & 7.01$\pm$1.7    & 4.21$\pm$1.0 && 3.05$\pm$0.5 & 6.6$\pm$0.5  \\
Centaurus & 156.2336 & -11.5868  & 38.7 & 10.8$\pm$3.9& 6.48$\pm$2.3& &1.51$\pm$0.4 & 7.58$\pm$1.3 \\
Hydra  & 139.4478 &	 -37.6063 & 41.0  & 4.39$\pm$1.3     & 2.63$\pm$0.8	&& 1.55$\pm$0.3 & 2.50$\pm$0.7\\
Perseus  &347.7159 	& -14.0594 &  52.8 &	16.3$\pm$4.2   & 9.78$\pm$2.5&& 2.32$\pm$1.6 & 2.46$\pm$1.8\\
Coma & 89.6226 & 8.1461 & 73.3  &15.9$\pm$1.2            & 9.54$\pm$0.7&&1.78$\pm$0.5 & 2.2$\pm$0.9	\\
\hline
\hline
\end{tabular}
\end{center}
\vspace{-0.25cm}
\caption{Clusters from \citet{2015AJ....149..171T} with H$_0$=75~km~s$^{-1}$~Mpc$^{-1}$: (1) cluster name, (2) supergalactic longitude, (3) supergalactic latitude, (4) distance, (5) virial mass, (6) virial mass converted. Simulated clusters from this paper: (7) M$_{200}$ for Tully grouping, (8) M$_{200}$ for Tempel grouping.}
\label{Tbl:1}
\end{table*}

In this section, the dark matter halos counterparts of local observed clusters are looked for and studied in the constrained simulations obtained with the two different grouping schemes. Five local clusters of different masses and at various distances from us are selected for further studies: Virgo, Coma, Perseus, Centaurus and Hydra. Their unique counterpart in each one of the constrained simulations is searched for in the list of dark matter halos extracted from the simulations with the halo finder. Note that only simulacra with masses higher than 10$^{14}\hmsun$ are considered. In addition, distances between simulacra and observed clusters cannot exceed 30\% of the distance estimate of the clusters. Regardless, the most important point is that if simulacra are slightly shifted in positions with respect to the observed cluster, their shifts are consistent so that their locations do not differ significantly from each other as shown hereafter.

\begin{figure*}
\includegraphics[width=0.91 \textwidth]{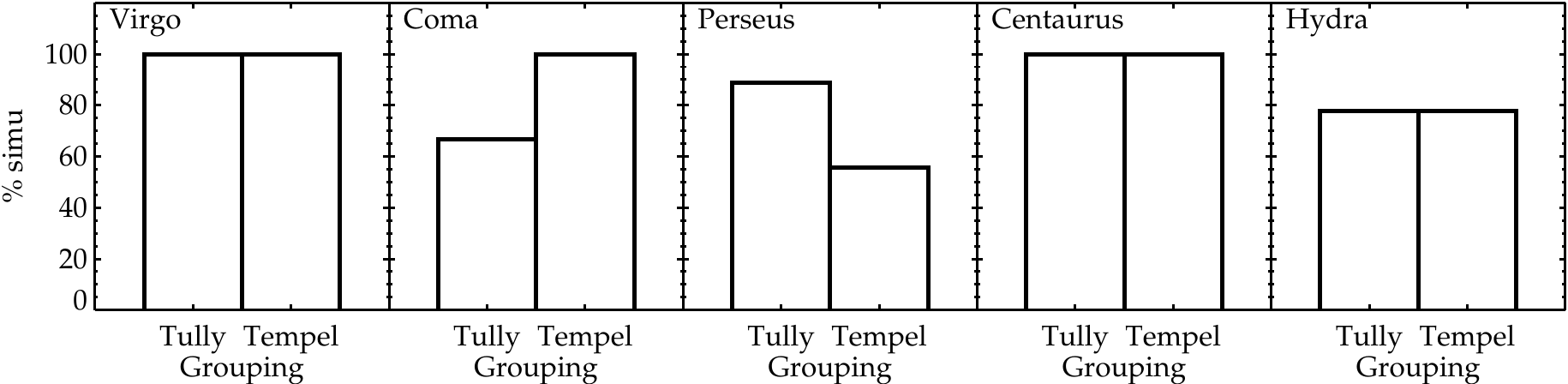} \\
\vspace{0.09cm}
\includegraphics[width=0.91 \textwidth]{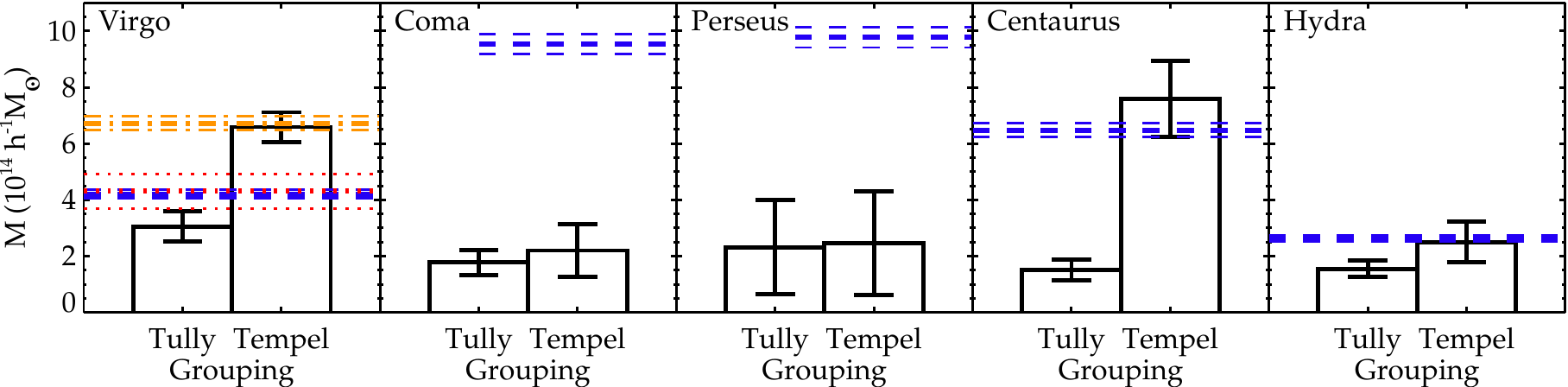} 
\caption{Top: percentage of simulations with a simulacrum of the observed cluster whose name is given in the top right corner of the panel for both type of grouping schemes. Bottom: average mass (histogram), standard deviation (error bar) of the different simulacra for both type of grouping schemes. Clearly the average masses of the different simulacra are higher when using Tempel grouping scheme. Blue thick dashed lines show the virial mass estimates from \citet{2015AJ....149..171T} converted to M$_{200}$. The blue thinned dashed lines show the 1$\sigma$ uncertainty in the conversion from M$_{\rm vir}$ to M$_{200}$. The red thick (thin) dotted lines show the latest observational mass estimate of the Virgo cluster from the first turn around radius converted to M$_{200}$ ($\pm$ 1$\sigma$ uncertainty in the conversion) while the orange dot-dashed lines stand for a dynamical mass estimate - equivalent to a virial mass for unrelaxed clusters - obtained studying galaxies in filaments falling into the Virgo cluster.}
\label{fig:clusters}
\end{figure*}

Fig. \ref{fig:clusters} gives the percentage of simulations in which a simulacrum of the observed clusters is found (top) as well as the average mass of the simulacra (bottom). Overall Tempel scheme does not increase significantly the percentage of success in getting a simulacrum: 87$\pm$14 against 89$\pm$19\%. This is in agreement with the fact that the local Large Scale Structure is well simulated in both cases, namely there are overdensity regions at the location of clusters. However, Tempel scheme increases the average mass of all the simulacra especially those of Virgo and Centaurus. While the mass of Virgo candidates is doubled (factor 2.2) that of Centaurus candidates is more than fivefold (factor 5.1) to reach an excellent agreement with recent observational estimates (within 2 and 1-$\sigma$ respectively). Indeed \citet{2015AJ....149..171T} published recently the virial masses of these local clusters in M$_\odot$ with distances consistent with H$_0$=75~km~s$^{-1}$~Mpc$^{-1}$. The only uncertainties related to the virial masses that are provided are those of the bi-weight projected virial radii. A propagation of uncertainty using the sole bi-weight project virial radii is far from optimal. Since \citet{2015AJ....149..171T} also supplies us with the luminosity masses that follow a 1:1 relation for clusters more massive than 10$^{14}~\hmsun$, the difference between the luminosity and the virial masses gives a rough estimate of the virial mass uncertainty. Table \ref{Tbl:1} summarizes these masses to be compared with M$_{200}$ (i.e. the mass enclosed in a sphere with a mean density of 200 times the critical density of the Universe) also included in the table. This mass derived by the halo finder is known to be proportional to the virial mass (given by the halo finder) via a factor of 0.80$\pm$0.03 \citep[e.g.][]{2016MNRAS.460.2015S}. Assuming the virial masses given by both observational estimates and the halo finder to be roughly similar, we overplot them for comparisons on Fig. \ref{fig:clusters} with blue thick dashed lines as well as the 1$\sigma$ uncertainty of the conversion with thinner lines. The orange dot-dashed lines represent a dynamical mass estimate of the Virgo cluster \citep{2015ApJ...807..122L} that can be assumed to be roughly similar to a virial mass estimate for an unrelaxed cluster. Although this value is higher than the general values found in the literature for the mass of the Virgo cluster \citep[e.g.][to give another reference]{2014ApJ...782....4K}, it is interesting to mention this value since it is based on the reconstruction of the dynamics of galaxies in filaments around the Virgo cluster. This value is thus obtained via both observation and numerical reconstructions like our value. The red dotted lines stand for a very recent estimate of Virgo's mass via the first turn around radius by \citet{2017arXiv171008935S} with M$_{200}$=4.9$\pm$0.7~M$_\odot$ \citep{2016MNRAS.460.2015S}. The average mass of the first Virgo clusters is within 2$\sigma$ of this estimate validating our previous study of the Virgo cluster with the constrained simulations that stated the good quality of the simulacra. Centaurus simulacra are now in extraordinary agreement with observations: their average mass is within 1$\sigma$ of the estimated mass. Hydra simulacra have now masses in excellent agreement with observational estimates as well: in Table \ref{Tbl:1}, the means are quasi-identical (2.6 and 2.5~$\times$10$^{14}~\hmsun$) and the standard deviation of the simulated halo masses is almost equal to the uncertainty of the observational mass estimate (0.8 and 0.7 $\times$10$^{14}~\hmsun$). While the mass of the Coma cluster is only increased by 24\% for Tempel scheme with respect to Tully's, it is worth noticing that it is now present in 100\% of the simulations. The only cluster that has less efficient simulacra is Perseus. However Perseus-Pisces region is poorly constrained with the second catalog of Cosmicflows. The newly released third catalog \citep{2016AJ....152...50T} that contains more constraints in that region allows us to foresee good simulacra for Perseus in a near future.\\

It is important to note that it is the combination of the grouping scheme and the \emph{bias minimization} that allows to get such results. Without the bias minimization scheme, Virgo's success rate drops to 60\% with masses barely above 10$^{14}~\hmsun$ as for Centaurus although its success rate is 100\%, the most massive simulacrum is about 4$\times$10$^{14}~\hmsun$ and the majority of the simulacra have masses barely above 10$^{14}~\hmsun$. \\

\begin{figure*}
\includegraphics[width=0.9 \textwidth]{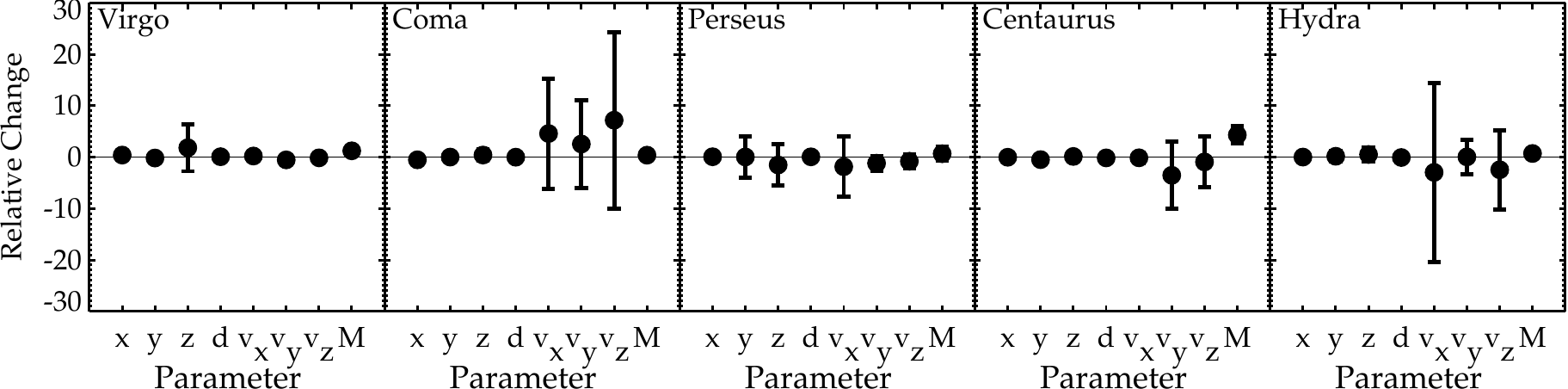} \\
\vspace{0.1cm}
\includegraphics[width=0.9 \textwidth]{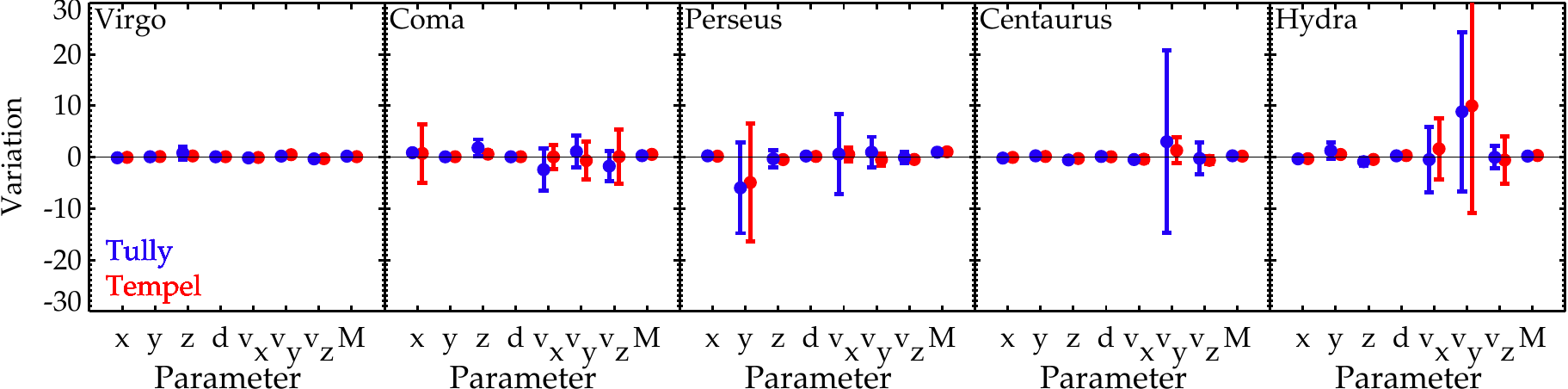} 
\caption{Top: relative change and standard deviation (filled circle and error bar)  between the parameters of the cluster simulacra obtained in the simulations produced with the two different grouping schemes. Bottom:  variation (filled circle) and standard deviation (error bar) of the parameters of the cluster simulacra found in the simulations obtained with Tully (blue) and Tempel (red) grouping schemes.}
\label{fig:clusprop}
\end{figure*}

The top of Fig. \ref{fig:clusprop} shows the relative change between the properties of the simulacra obtained in the two sets of constrained simulations. We define the relative change as the difference between the parameter value of the dark matter halo in Tempel constrained simulation and that in Tully's divided by the value in Tully's. Interestingly Virgo already showed to be very stable \citep{2016MNRAS.460.2015S} is quasi unchanged in Tempel constrained simulations with respect to Tully's. The only exception is the \textit{z} supergalactic coordinate. This is expected as the \textit{z} direction is the less constrained because of the zone of avoidance. The other clusters present simulacra those relative changes are the most important for the velocity components. Perseus presents also relative changes of 5 to 10 times the \textit{y} and \textit{z} supergalactic coordinates confirming that it is the less constrained clusters in terms of positions. Still all the maximum relative changes are held below 10 times the parameters.

The bottom of Fig. \ref{fig:clusprop} gives the mean variation (filled circle) and standard deviation (error bar) of the properties of the dark matter halos in Tempel (red) and Tully (blue) constrained simulations taken separately. The variation is defined as the standard deviation of a parameter divided by the parameter value of a given simulacrum. Virgo appears very stable for both grouping schemes and even more stable (by a factor 5) in terms of the \textit{z} supergalactic coordinate when using Tempel grouping scheme rather than Tully's. The \textit{y} and \textit{z} velocity components of Centaurus counterparts are more stable 
when using Tempel grouping scheme. For the \textit{y} component, the standard deviation of the variation decreases from 18\% to less than 3\%. The other clusters have more mitigated variation and standard deviation values.  \\

Overall the efficiency of changing the grouping scheme for a less aggressive scheme \citep[see][for a detailed discussion]{2017MNRAS.469.2859S} results in galaxy clusters more:
\begin{enumerate}
\item present: on average clusters are simulated with a success rate increased by 5\% with respect to Tully grouping scheme;
\item stable: on average the stability of the parameters is increased by a factor 3 with respect to the parameters of halos obtained with Tully grouping scheme;
\item massive: on average halos are 39\% more massive than those obtained with Tully grouping scheme.
\end{enumerate}
These assertions are reinforced when considering Centaurus and Virgo: Centaurus simulated mass is now within 1-$\sigma$ of the observation mass estimate. Virgo is perfected: its z-component is in particular more constrained than before with a standard deviation decreased by a factor 5.
 
\section{Conclusion}
Galaxy clusters are excellent cosmological probes whose formation and evolution still need to be understand in detail. Numerical simulations of clusters constitute a formidable complementary approach to their observations. However, the diversity of galaxy clusters complicates comparisons with their numerical counterparts on a one-to-one basis down to the simulated and observed galaxy populations.

Such detailed comparisons are feasible in the context of simulations that resemble the local Universe provided that the latter reproduce the local Large Scale Structure down to the cluster scales. In our first generation of constrained simulations made with the second catalog of galaxy peculiar velocity of the Cosmicflows project combined with a bias minimization scheme, large overdensities were present at the location of local clusters but massive enough dark matter simulacra of the latter were found only for the Virgo cluster. 

In a previous study, we showed that one of the essential step in the process of building the constrained initial conditions affects the overdensity values of the reconstructed field. This step consists in grouping the catalog of constraints (galaxies and their peculiar velocities) to remove non-linear motions that would affect the linear reconstruction. However, this first study demonstrated that the grouping must be made with parsimony to preserve the infall on clusters and thus to increase the local densities. This study goes further as it probes the impact of the grouping scheme (Tully's based on the literature and Tempel's based on an advanced FoF algorithm) on the final product: the simulations that resemble the local Universe.

Overall, the same Large Scale Structure is simulated with both grouping schemes. However, a slight increase (10\%) of the power spectrum on large scales is observed with the less aggressive grouping scheme. This is ideal as it improves the probability of the power spectrum of the local Universe given Planck power spectrum. Tempel grouping scheme also increases the mass function at the high end, more precisely the most massive halos are heavier than with Tully grouping scheme. These new observations imply the real need to inquire further on the impact of the data modeling on the resulting simulations. It is essential to determine whether the residual is intrinsic to the data or due to some data modeling or both. This study has shown that the grouping is partly responsible, in an ongoing study we are investigating the impact of the data uncertainty modeling.

A thorough study of 5 of the local clusters (Virgo, Centaurus, Hydra, Coma and Perseus) reveals that their simulacra are better representative. The Virgo simulacrum is still very stable (present in 100\% of the simulations) and its mass is increased by 50\% with respect to the first generation of constrained simulations we produced. In both cases, the masses are within 2-$\sigma$ of the observational mass estimate, the previous one on the low side, the new one on the upper side. The \textit{z} supergalactic coordinate of the simulacra presents a standard deviation divided by a factor 5, implying that although the \textit{z} direction is that of the zone of avoidance, it is possible to constrain further the \textit{z} position of the Virgo dark matter halos with a moderate grouping scheme.
The most incredible advantage of Tempel grouping scheme is visible for the Centaurus cluster. In the new set of constrained simulations, Centaurus simulacra are five times more massive than before and are within 1-$\sigma$ of the recent observational estimates. Coma is also improved in the sense that a simulacrum is now present in 100\% of the simulation with a mass increased by nearly 25\%. Hydra's and Perseus simulacra are also slightly more massive than in the first generation of simulations. The formers, present in 80\% of the simulations, have the quasi same mean (within 5\%) and standard deviation (within 12\%) as the observational estimate.

All in all, using Tempel grouping scheme improves considerably the simulacra of Centaurus and perfects those of Virgo \emph{provided that it is combined with the bias minimization scheme}. Indeed, without the latter all the advantages of using Tempel rather than Tully grouping scheme disappear. The combination of Tempel grouping scheme and the bias minimization scheme ameliorates the simulacra of Coma, Hydra and Perseus although there is still room for improvements. First the third catalog of peculiar velocities of the Cosmicflows project will offer us more data especially in the direction of Perseus-Pisces. Second a better modeling of the uncertainties in the bias minimization scheme (so far a 5\% uncertainty is applied to all the distances obtained after minimization and is propagated to the velocities) is under study. Third a new grouping algorithm based on point processes with interactions is investigated.

Now the constrained simulations of the local Universe produced via the method described in this paper that still uses \emph{only peculiar velocity datasets as constraints} (in the sense that no additional density constraints are added at the positions of the clusters, the velocity-constraints contain both the position and mass information of the clusters) not only resemble the local Large Scale Structure and have Virgo dark matter simulacra but also stable Centaurus halos with masses within 1-$\sigma$ of observational estimates as well as better representatives of Coma, Hydra and Perseus.

A large number of zoom-in dark matter simulations of these halos will permit making statistical studies of these local clusters regarding their formation, their substructures, etc. In addition, further zoom-in hydrodynamical simulations of these halos are planned to study the galaxy populations of these various local clusters to be compared with their observational counterparts. Links between properties of galaxy populations in local clusters of different types (various masses, formation histories, substructures, etc) will be highlighted to further refine our understanding of galaxy formation and evolution in clusters.

\section*{Acknowledgements}
The authors would like to thank the referee, Francisco-Shu Kitaura, and Stefan Gottl\"ober for providing comments that helped clarified the paper as well as Sergey Pilipenko for very useful discussions. JS acknowledges support from the Astronomy ESFRI and Research Infrastructure Cluster ASTERICS project, funded by the European Commission under the Horizon 2020 Programme (GA 653477) as well as from the ``l'Or\'eal-UNESCO Pour les femmes et la Science'' and the ``Centre National d'\'etudes spatiales (CNES)'' postdoctoral fellowship programs. ET was supported by ETAg grants IUT40-2, IUT26-2 and by EU through the ERDF CoE grant TK133. The authors gratefully acknowledge the Gauss Centre for Supercomputing e.V. (www.gauss-centre.eu) for providing computing time on the GCS Supercomputers SuperMUC at LRZ Munich and Jureca at JSC Juelich.  


\bibliographystyle{mnras}

\bibliography{biblicomplete}
 \label{lastpage}
\end{document}